\documentclass[aps,prb,twocolumn,superscriptaddress,showpacs]{%
revtex4}

 \usepackage{graphicx}

\newcommand{\iside}{\ensuremath{I_{\text{side}}{}}}
\newcommand{\isidemin}{\ensuremath{I_{\text{side}}^{\text{min}}{}}}
\newcommand{\vside}{\ensuremath{V_{\text{side}}{}}}
\newcommand{\iC}{\ensuremath{I_{\text{C}}}}
\newcommand{\iE}{\ensuremath{I_{\text{E}}}}
\newcommand{\vBC}{\ensuremath{V_{\text{BC}}}}
\newcommand{\vBE}{\ensuremath{V_{\text{BE}}}}
\newcommand{\vE}{\ensuremath{V_{\text{E}}}}
\newcommand{\eBC}{\ensuremath{E_{\text{BC}}}}
\newcommand{\eF}{\ensuremath{E_{\text{F}}}}
\newcommand{\lEC}{\ensuremath{L_{\text{EC}}}}
\newcommand{\lee}{\ensuremath{l_{\text{ee}}}}
\newcommand{\ekin}{\ensuremath{E_{\text{kin}}}}
\newcommand{\amax}{\ensuremath{\alpha_{\text{max}}}}

\usepackage[normalem]{ulem}

\usepackage{color}

\begin{document}

\title{Electron-avalanche amplifier based on the electronic Venturi effect}



\author{D. Taubert}
\affiliation{Center for NanoScience and Fakult\"at f\"ur Physik,
Ludwig-Maximilians-Universit\"at,
Geschwister-Scholl-Platz 1, 80539 M\"unchen, Germany}

\author{G. J. Schinner}
\affiliation{Center for NanoScience and Fakult\"at f\"ur Physik,
Ludwig-Maximilians-Universit\"at,
Geschwister-Scholl-Platz 1, 80539 M\"unchen, Germany}

\author{H. P. Tranitz}
\affiliation{Institut f\"ur Experimentelle Physik, Universit\"at Regensburg,
93040 Regensburg, Germany}

\author{W. Wegscheider}
\affiliation{Solid State Physics Laboratory, ETH Zurich, 8093 Zurich,
Switzerland}

\author{C. Tomaras}
\affiliation{Center for NanoScience and Fakult\"at f\"ur Physik,
Ludwig-Maximilians-Universit\"at,
Geschwister-Scholl-Platz 1, 80539 M\"unchen, Germany}

\author{S. Kehrein}
\affiliation{Center for NanoScience and Fakult\"at f\"ur Physik,
Ludwig-Maximilians-Universit\"at,
Geschwister-Scholl-Platz 1, 80539 M\"unchen, Germany}

\author{S. Ludwig}
\affiliation{Center for NanoScience and Fakult\"at f\"ur Physik,
Ludwig-Maximilians-Universit\"at,
Geschwister-Scholl-Platz 1, 80539 M\"unchen, Germany}

\date{\today}

\begin{abstract}
Scattering of otherwise ballistic electrons far from equilibrium is investigated in a cold
two-dimensional electron system. The interaction between excited electrons and the
degenerate Fermi liquid induces a positive charge in a nanoscale region which would be
negatively charged for diffusive transport at local thermal equilibrium. In a
three-terminal device we observe avalanche amplification of electrical current, resulting
in a situation comparable to the Venturi effect in hydrodynamics. Numerical calculations
using a random-phase approximation are in agreement with our data and suggest Coulomb
interaction as the dominant scattering mechanism.
\vspace*{-8mm}
\end{abstract}

\maketitle



\begin{figure}
\includegraphics[width=\columnwidth]{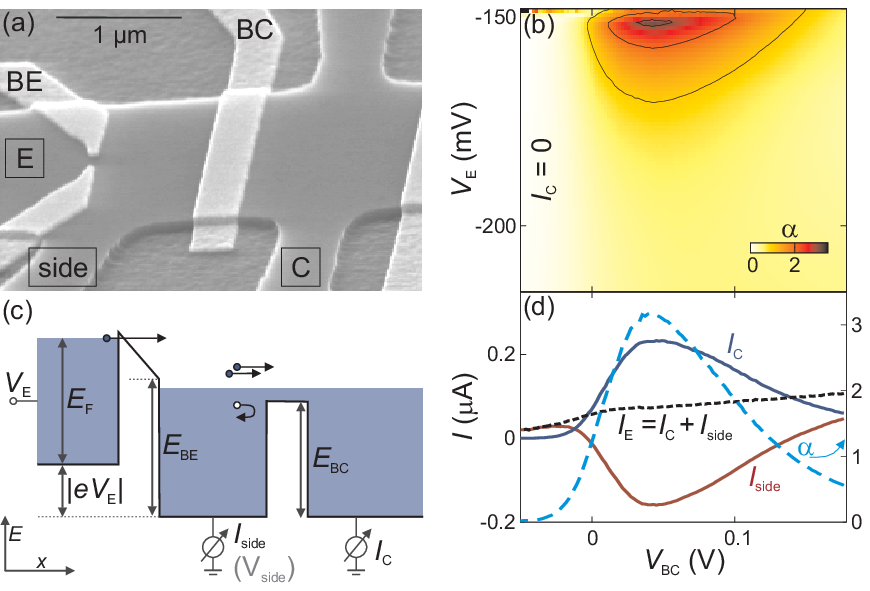}
\caption{\label{fig1}(Color online) (a) Electron micrograph of the hallbar
(elevated area), fabricated by wet etching, which contains the 2DES. Top gates
(light gray) are
used to define electrostatic barriers BE and BC. Ohmic contacts E, side, and C
are marked. (b) Transfer ratio $\alpha = \iC / \iE$,
plotted against barrier voltage \vBC\ and bias \vE\ ($\vBE =-0.925\,\mbox{V}$).
Contour
lines show $\alpha=1,2,3$. (c) Energy diagram sketching the experiment. (d)
Relevant currents as well as transfer ratio along a horizontal cross-section of
Fig.\,\ref{fig1}(b) at $\vE = -153 \mbox{ mV}$.}
\vspace{-5mm}
\end{figure}

Bernoulli's principle states that an increase in velocity of an inviscid fluid is
accompanied by a pressure decrease. A related ``hydrodynamic'' effect on the nanoscale has
been predicted by Govorov \emph{et al.}\ \cite{sasha} who consider a degenerate high-mobility
Fermi liquid instead of a classical inviscid fluid. \emph{Hot} electrons are injected
through a quantum point contact (QPC) and then move ballistically along a two-dimensional
electron system (2DES). They transfer energy and forward momentum to electrons from the
degenerate Fermi sea which causes a net positive charge to be left behind. This effect,
based on momentum transfer, has so far eluded experimental proof. In classical
hydrodynamics Bernoulli's principle combined with the continuity equation leads to the
Venturi effect. That is, the pressure in a fluid decreases as it passes through a tube
with reduced cross section. In a spectacular application, the water jet pump introduced by
Bunsen in 1869,\cite{bunsen} the reduced pressure is utilized for evacuating a side port.
After passing the side port the fluid is decelerated into a wider collector tube which
also seals the pump from its exhaust and improves the vacuum.  Here we present a nanoscale
device which behaves similarly to a water jet pump, ``pumping'' electrons instead of a
classical fluid. Our electron jet pump follows the idea described in Ref.\ \onlinecite{sasha} but is
enhanced by an additional barrier ``BC'' that separates the side contact from
the collector [see Fig.\,\ref{fig1}(a); electrons are injected from the left].
Excited electrons which carry enough forward momentum can pass BC and
reach the collector contact ``C'' but positively charged holes (in the Fermi
sea) are reflected. If the
side contact is grounded, the positive charge is neutralized by electrons
flowing from the
side into the device. This flow adds to the electron current from the emitter to create an
amplified current at the collector port.  Our electron jet pump is therefore a prototype
of a ballistic electron-avalanche amplifier. We observe amplification up to a factor of
seven which hints at several electron-electron scattering events per electron between
emitter (QPC) and collector.

Amplifiers based on the injection of \emph{hot} electrons have been pursued
since the 1980s in various transistor structures
\cite{heiblum_theta,brill,kaya1996} and high-mobility 2DESs 
\cite{sivan1989,palevski,kaya}. Our systematic investigations go well beyond
those previous publications and give the perspective of a detailed
understanding of nonequilibrium transport in Fermi liquids. While this also
includes the emission of acoustic \cite{georg} and optical
\cite{sivan1989,dzurak} phonons and plasmons, here we focus on scattering
between electrons (see e.g. Ref.\ \onlinecite{buhmann} which considers much lower energies
than covered here). In our experiments we realize a transition from a regime in
which the electron-electron scattering length \lee\ is small compared to the
sample dimensions (avalanche amplification) to purely ballistic motion of hot electrons. Our avalanche amplifier also
promises future applications, e.\,g.\ as a new kind of charge detector.

The device shown in Fig.\,\ref{fig1}(a)
has been fabricated from a GaAs/AlGaAs heterostructure which contains a 2DES
90\,nm
below the surface. The sample consists of a hallbar with several ohmic contacts
(not visible). Three
terminals are used as emitter ``E,'' ``side'' contact, and collector C.
Metallic gates (light
gray) serve to define a broad collector barrier BC and an emitter QPC ``BE''
electrostatically.
The use of a QPC as emitter is not crucial; very similar data have been taken
with a broad
emitter barrier instead. As sketched in Fig.\,\ref{fig1}(c), electrons are
injected from the emitter
E at potential $\vE < 0$ into the region between BE and BC. By tuning BE near
pinch-off, it is
assured that the injected electrons have a kinetic energy close to $\left| e \vE
\right|+\eF$ (\eF\ is the
Fermi energy). At first these \emph{hot} electrons move ballistically towards BC.
Eventually they
scatter and excite additional electrons from the degenerate Fermi sea, thereby
transferring part of
their energy and momentum. Conduction-band holes in the Fermi sea are left
behind [Fig.\,\ref{fig1}(c)]. The collector
barrier separates excited electrons (which can pass the barrier) from the holes
(which are
reflected); the accumulation of holes causes a buildup of positive charge
between BE and BC. The
measured currents \iC\ and \iside\ are defined to be positive when electrons
flow from the sample \emph{into} the respective
terminals, as would be expected in diffusive transport. Here, we tune our devices away from the diffusive-transport
regime. In contrast to many previous publications \cite{heiblum_theta, brill,kaya1996} we reach a ballistic regime
which is far from local thermal equilibrium.

The mobility and Fermi energy of the 2DES are $\mu=1.4 \times
10^6\,\text{cm}^2/\text{Vs}$ (at $ T
\approx 1\,\text{K}$) and $\eF = 9.7\,\text{meV}$ (carrier density $n_s=2.7
\times
10^{15}\,\text{m}^{-2}$). In our case the elastic mean-free path
$l_{\text{m}}\simeq12\,\mu$m exceeds the dimensions of the nanostructure by far. Measurements shown here have been performed in a $^3$He cryostat at
$T_{\text{bath}}\simeq260\,$mK but comparable results have been obtained in a
wide temperature
range of $20\,\text{mK}\le T_{\text{bath}}\le 20\,$K in similar samples.

To probe for amplification, we consider the transfer ratio $\alpha = \iC / \iE$ with $\iE\equiv\iC+\iside$ the current
injected
from E. As a typical example $\alpha$ is plotted in Fig.\,\ref{fig1}(b) as a
function of \vE\ and $\vBC$.
Amplification ($\alpha>1$) is observed in a limited region which is framed by
contour lines.
We have already reached $\alpha\simeq7$ in a similar setup (here 
$\alpha\lesssim3.2$). The actually
measured  currents \iC\ and \iside\ are shown in Fig.\,\ref{fig1}(d) for constant
$\vE=-153\,$mV.
For very negative \vBC, the collector barrier BC is closed, $\iside=\iE$, and
$\alpha=0=\iC$. As BC
is opened, \iC\ shows a broad maximum, exceeding \iE. Hence, electrons are drawn
in from the side
contact ($\iside<0$, $\alpha>1$), making the device an electron jet pump. In the
limit of a wide-open
collector barrier (large \vBC) the electron-hole selectivity is lost and the
setup behaves similarly to a network of ohmic resistors.

\begin{figure}
\includegraphics[width=\columnwidth]{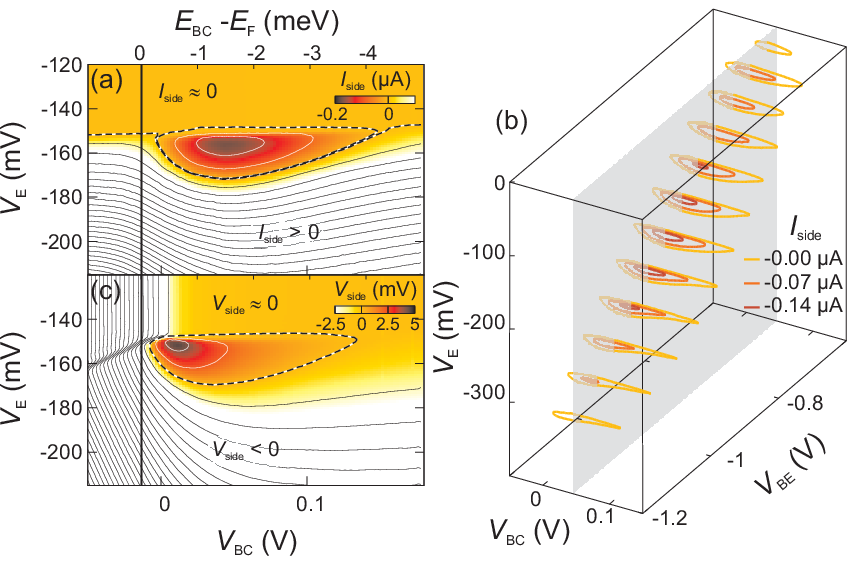}
\caption{\label{fig2}(Color online) (a) \iside\ as a function of \vE\ and \vBC\ [same raw data
as in Fig.\,\ref{fig1}(b)]. Contour lines are spaced by
$0.07\,\mbox{$\mu$A}$ (white for $\iside<0$, dashed for $\iside\simeq0$).
(b) 3D representation of $\iside\le0$ (three contour lines) vs \vE\ and \vBC\
for several \vBE.
(c) \vside\ measured at floating side contact, parameters as in
Fig.\,\ref{fig2}(a).
Contour lines are spaced by 2\,mV (white for
$\vside>0$, dashed for $\vside\simeq0$).}
\vspace{-5mm}
\end{figure}

Figure \ref{fig2}(a)
shows \iside\ as a function of \vE\ and \vBC\ [same raw data as
Fig.\,\ref{fig1}(b)]. For $\vE \gtrsim
-150\,\mbox{meV}$ BE is completely closed, hence, current flow is
suppressed ($\iside =\iC =\iE =0$). As \vE\ is
increased the emitter QPC opens and \iE\ becomes nonzero. A dashed
contour line encloses the region of $\iside<0$. The three-dimensional (3D) representation in Fig.\,\ref{fig2}(b)
displays a few contour lines at
$\iside\le0$ as a function of \vE\ and \vBC\ for several emitter configurations
\vBE. Clearly $\iside<0$ only
occurs within a narrow tube in a region where the emitter QPC BE is almost
pinched off. 

The dependence of $\alpha$ on the collector barrier
height \eBC\ is also shown in Fig.\,\ref{fig2}(a) (top axis). \eBC\ can be determined from \vBC\
by measuring the reflection of Landau levels on the barrier in a magnetic field.\cite{komiyama,haug} In addition, the
calibration point $\eBC = \eF$ is known from linear-response transport
measurements across the barrier as a function of \vBC. A simple one-dimensional (1D) model predicts maximal
amplification \amax\ at exactly $\eBC=\eF$ since in
this case excited electrons
would pass BC whereas holes would be reflected. Strikingly, in Fig.\ \ref{fig2}(a) \amax\ (which almost
coincides with \isidemin)
occurs at $\eBC<\eF$ ($\eBC\simeq\eF-1.4\,\mbox{meV}$). This is related to the 
2D character of the charge carriers which allows an angle distribution of
momenta within the 2DES. A charge carrier can only pass BC if its forward momentum component $p_\perp$ perpendicular
to
the barrier fulfills $p_\perp^2/2m > \eBC$. Compared to the 1D case the barrier has therefore to be lower in 2D for
a significant portion of the excited electrons to pass. \amax\ is thus expected at $\eBC<\eF$,\cite{angle_effect} which
is in agreement with
experimental data (\emph{angle effect}). In a previous publication, \amax\ at  $\eBC>\eF$ was reported,\cite{kaya} 
though this was obtained with a very different calibration procedure.

As an alternative to measuring \iside\ in a three-terminal setup, Fig.\,\ref{fig2}(c) shows \vside\
detected at the floating side contact. In the diffusive regime $\vside<0$ would be expected (since $\vE<0$).
However, as in the current measurement, scattering of the injected electrons occurs and causes $\vside>0$ in a region
roughly
comparable to that of $\iside<0$ in Fig.\,\ref{fig2}(a). Since this is a two-terminal setup [see sketch in
Fig.\,\ref{fig1}(c)], the continuity equation forces $\iC = \iE$. Electrons
cannot escape to the side contact and, hence, the \emph{angle effect} as described above must be absent. Nevertheless
the maximum of $\vside>0$ is still observed at $\eBC<\eF$. This can be explained by means of a positive charge that
builds up between BE und BC in steady state (in a
current measurement this charge is at least partly neutralized by $\iside < 0$). The
positive charge causes a decrease in the local chemical potential. Electrons trying to escape via BC thus see a larger
\emph{effective} barrier, and hence, the maximum effect is again found at $\eBC<\eF$ (\emph{charge
effect}).\cite{charge_effect}

\begin{figure}
\includegraphics[width=\columnwidth]{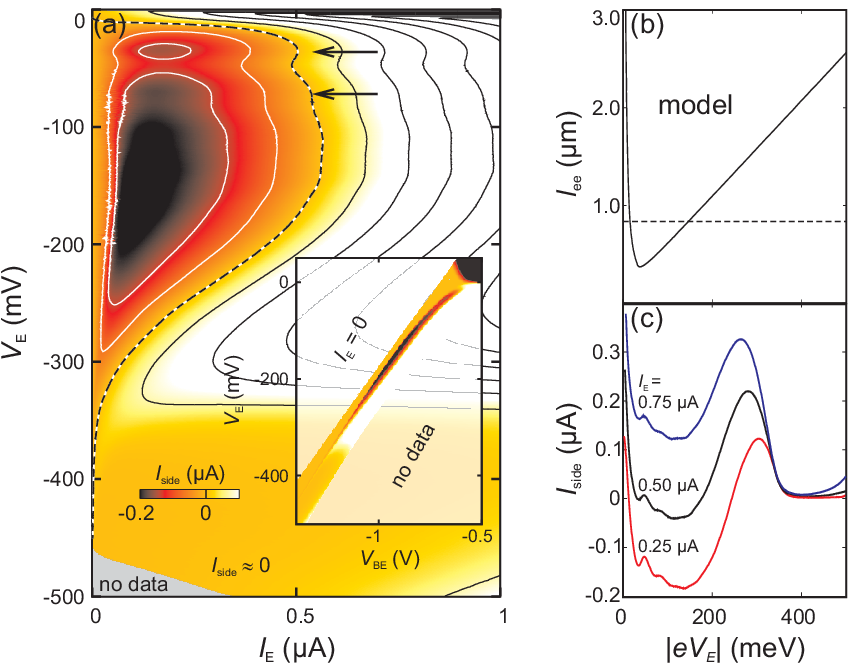}
\caption{\label{fig3}(Color online) (a) Inset: \iside\ as a function of \vBE\ and \vE. No data
exist for $\vE\times\iside>700\,$nW and
in the upper left region where the emitter QPC is closed and all currents
vanish.
Main plot: same data as a function of injected current \iE. 
(b) Numerical calculation of the electron-electron scattering length, \lee, as a
function of excess kinetic 
energy $|e\vE| \simeq E_{\text{kin}}-\eF$ at $T=0$.
(c) Vertical slices of Fig.\,\ref{fig3}(a) for three different \iE.}
\vspace{-5mm}
\end{figure}

In Fig.\,\ref{fig2}(b) a semi-transparent plane perpendicular to the cross section
of
Fig.\,\ref{fig2}(a) marks constant $\vBC = 0.045\,\mbox{V}$. A detailed
measurement of \iside\ within
this plane is plotted in the inset of
Fig.\,\ref{fig3}(a).
The main plot shows the same \iside\ data as a function of \vE\ and the overall
current \iE. We observe $\iside<0$ to
the left of the dashed contour line. Figure \ref{fig3}(c) displays vertical
cross sections of Fig.\,\ref{fig3}(a) which show the
dependence of \iside\ on the energy of the injected electrons for constant \iE.
The main behavior seen in
Figs.\,\ref{fig3}(a) and \ref{fig3}(c) is closely related to the electron-electron
scattering length \lee. By numerical calculations based on the random-phase
approximation, we have
extended predictions for the linear-response regime \cite{chaplik,giuliani} to
the nonequilibrium
case required for our experiments. The calculated \lee\
is plotted in Fig.\,\ref{fig3}(b) for $T=0$ as a function of $|e\vE| \simeq
\ekin-\eF$, the excess kinetic energy of the
injected electrons. As expected, \lee\ diverges for $\ekin\to\eF$ ($e\vE\to0$)
and then rapidly decreases as \ekin\ is
increased.\cite{chaplik,giuliani} At higher \ekin, a minimum at
$|e\vE|\simeq5\eF$ is followed by a linear increase in \lee\ which can be
understood in terms of
decreasing electron-electron interaction times.

This behavior can be mapped onto the measured energy dependence of \iside\
[Fig.\,\ref{fig3}(c)] by
taking into account the sample geometry [Fig.\,\ref{fig1}(a)]. The distance
between BE
and BC is $\lEC\simeq840\,$nm [dashed line in Fig.\,\ref{fig3}(b)], about twice
as long as the minimal calculated \lee. In the extreme limits of $\ekin\to\eF$
or $\ekin\to\infty$,
we find $\lee\gg\lEC$ and expect electrons to move ballistically and without
electron-electron scattering within the sample. As the energy is
increased starting from $e\vE=0$, \lee\ decreases, and for $\lee<\lEC$, a
positive charge builds up
between BE and BC. It is neutralized by a growing negative component of \iside\
[Fig.\,\ref{fig3}(c)].
Excited electrons always lose energy when scattering with the cold Fermi sea.
Hence, scattering of
an excited electron on the negative slope of $\lee(e\vE)$ [Fig.\,\ref{fig3}(b)]
results in carriers with
increased \lee\ for subsequent scattering events. In contrast, scattering of
electrons on the positive slope of $\lee(e\vE)$ often results in carriers with decreased \lee. These
carriers contribute
heavily to a negative \iside\ by multiple scattering events. The measured
$\isidemin(e\vE)$
clearly is expected to extend to higher energies compared to the minimum of
$\lee(e\vE)$ [Figs.\,\ref{fig3}(b)
and \ref{fig3}(c)]. At larger energies the injected electrons tend to pass BC and
scatter beyond the
barrier. For \lee\ only slightly larger than \lEC, some of the scattered
electrons can travel back across BC and into the side contact, causing the local
maximum of $\iside(e\vE)>0$ visible in Fig.\,\ref{fig3}(c). For even higher \ekin\
($e\vE>350\,$mV, $\lee\gtrsim3\lEC$) we find an extended regime with
$\iside\simeq0$ ($\iC\simeq\iE>0$). Here \lee{} exceeds the sample dimensions by
far so that electron-electron scattering happens far beyond BC, and all
resulting charge carriers end up in the grounded collector contact C. This behavior ($\iside\simeq0$) emphasizes
the ballistic nature of the
hot electrons in our experiments which goes beyond previously published results.\cite{heiblum_theta,brill,kaya1996}

\begin{figure}
\includegraphics[width=\columnwidth]{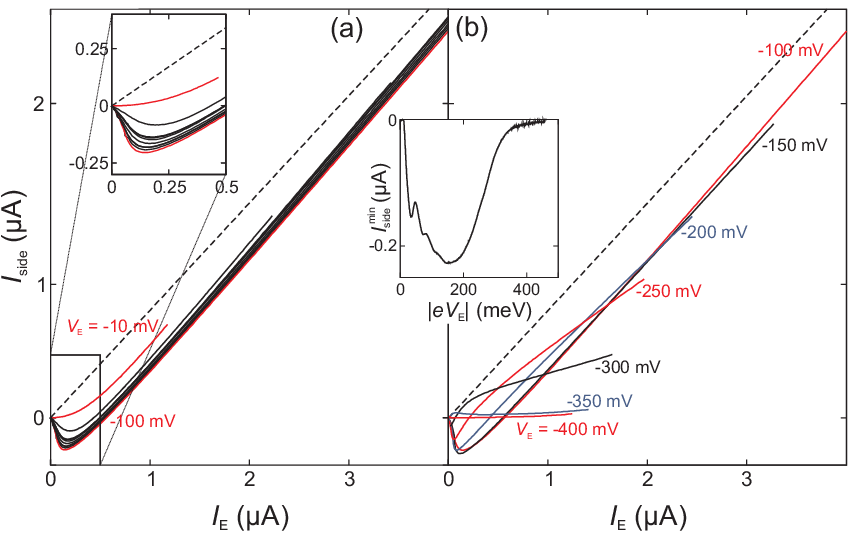}
\caption{\label{fig4}(Color online) \iside\ as a function of injected current \iE\ for
$\left| \vE \right| \le100\,$mV (spacing 10\,mV) in (a) and $\left |\vE \right|
\ge100\,$mV in
(b). Dashed straight lines represent ohmic behavior (see main text). The central
inset plots
\isidemin\ vs the energy of the injected electrons $|e\vE|$.}
\vspace{-5mm}
\end{figure}

Cross sections of Fig.\,\ref{fig3}(a) at constant excess kinetic energy $|e\vE|$
allow us to discuss the
dependence of \iside\ on the total current \iE and are displayed as line plots
for 
$\left| \vE \right| \le100\,$mV in Fig.\,\ref{fig4}(a)
and for $\left| \vE \right| \ge100\,$mV in Fig.\,\ref{fig4}(b). The dashed
straight lines
represent the expectation for ohmic behavior ($\iside\propto\iE$) with a slope
determined by
measuring the two-terminal resistances of the device in the linear-response
regime. For
$\iE\gtrsim0.3\,\mbox{$\mu$A}$ and  within the broad minimum of $\iside(e\vE)$
seen in
Fig.\,\ref{fig3}(c) ($40\,\mbox{meV}\lesssim \left| \vE \right| \lesssim150\,$mV),
all curves  coincide and are almost
parallel to the ohmic line. This behavior is plausible for small enough \lee\
when multiple
scattering events and reflections, e.\,g.\ at BC, scramble the electrons. Under
these conditions the
initial momentum of the injected electrons becomes less important for the
direction of current.
However, in our case ohmic behavior is superimposed with a ballistic effect, a
negative
contribution to \iside\ due to the separation of electrons and holes at BC. 

As
\iE\ is increased
from $\iE=0$ by adjusting \vBE, more and more electron-hole pairs are created
and partly separated
at BC. Only part of the positive charge of the holes can be neutralized from the
side contact due to its finite resistance. The remaining positive charges lower the chemical
potential of the Fermi
sea between BE and BC (\emph{charge effect}) so that less excited electrons
escape via BC. The
reflected electrons cause additional neutralization of holes. The neutralization
rate therefore
increases with increased \iE, and the steady-state negative component of \iside,
reached when hole
creation and neutralization rates balance, saturates for large \iE. For higher
energies and
longer \lee, deviations from this behavior occur as the injected electrons move
forward ballistically [Fig.\,\ref{fig4}(b)].

The central inset in Fig.\,\ref{fig4} displays \isidemin\ as a function of the
energy $|e\vE|$ of
the injected electrons. Similar to Fig.\,\ref{fig3}(c) it once again states the
strong energy 
dependence of the amplification effect already discussed above. The local minima
at
$|e\vE|\simeq36\,$meV and 72\,meV are caused by emission of optical phonons.
\cite{sivan1989,dzurak} They can also be seen in Fig.\,\ref{fig3}(c) and more
distinct in Fig.\,\ref{fig3}(a) (two black arrows at $\vE=36$ and 72\,meV).

Finally, it is important to differentiate the observed electronic Venturi effect from a
thermoelectric effect caused by Joule heating. Such thermal effects are usually described
within local equilibrium in the diffusive regime and mainly depend on the dissipated
power. Figure \ref{fig4} can be used to analyze \iside\ as a function of power $P = \left|
\vE \iE \right|$ since \vE\ is constant for each curve. The negative contribution to
\iside\  saturates as $P$ is increased, whereas in a thermoelectric effect it would be
expected to grow further. The amplification ratios for thermally driven effects are also
expected to be much smaller \cite{brill} compared to the $\alpha$ we find, which again
confirms the role of ballistic motion for the observed effects. In addition, the
strong dependence of $\alpha$ on the energy of the injected electrons
[Fig.\,\ref{fig3}(a)], as well as the maximum $\alpha$ occurring for $\eBC<\eF$ [Fig.\,\ref{fig2}(a)], are in direct
contradiction to an interpretation
in terms of a thermoelectric effect.

As a result, we have built a prototype of an electron-avalanche amplifier. It
is based on a jet pump explained by the electronic Venturi effect, namely,
scattering of \emph{hot} electrons with a degenerate Fermi liquid. Our systematic
investigations go well beyond
earlier publications and provide a comprehensive picture of the physics involved
in the ballistic
nonequilibrium regime. We present a consistent model based on electron-electron
scattering
and electron-hole neutralization which agrees qualitatively with our
experimental results.
Modifications in geometry and circuitry will result in improved electron jet
pumps with potential
applications, e.\,g., as a non-invasive charge detector. 
In such a device a single electron originating from a quantum dot would trigger
a current pulse 
strong enough to be detected.

We thank J.\,P.\ Kotthaus, A.\ Govorov, L.\ Molenkamp, M.\ Heiblum, I.\ Kaya, and F.\ Marquardt for fruitful
discussions. Financial support by the German Science Foundation via SFB 631, SFB 689, LU 819/4-1, and the German Israel
program DIP, the German Excellence Initiative via the ``Nanosystems Initiative Munich (NIM)," and LMUinnovativ (FuNS)
is gratefully acknowledged.


\begin{thebibliography}{13}
\bibitem{charge_effect}
\bibinfo{author}{In the regime where BE and BC are closed [top left corner of Fig.\,\ref{fig2}(c)], the side
contact is
electrically insulated. Here the potential \vside\ is frozen at a constant value depending on the experimental history
[in Fig.\ \ref{fig2}(c) we have performed vertical sweeps].}
\end{thebibliography}

\end{document}